# Efficient methodology for implementation of Encrypted File System in User Space

Dr. Shishir Kumar
Department of CSE, Jaypee
Institute of Engg. & Technology
Guna (M.P.), India
dr.shishir@yahoo.com

U.S. Rawat
Department of CSE, Jaypee
Institute of Engg. & Technology
Guna (M.P.), India
umasrawat@gmail.com

Sameer Kumar Jasra
Department of CSE, Jaypee
Institute of Engg. & Technology
Guna (M.P.), India

Akshay Kumar Jain
Department of CSE, Jaypee
Institute of Engg. & Technology
Guna (M.P.), India

*Abstract*— **The Encrypted File System (EFS) pushes encryption services into the file system itself. EFS supports secure storage at the system level through a standard UNIX file system interface to encrypted files. User can associate a cryptographic key with the directories they wish to protect. Files in these directories (as well as their pathname components) are transparently encrypted and decrypted with the specified key without further user intervention; clear text is never stored on a disk or sent to a remote file server. EFS can use any available file system for its underlying storage without modifications, including remote file servers such as NFS. System management functions, such as file backup, work in a normal manner and without knowledge of the key. Performance is an important factor to users since encryption can be time consuming.**

**This paper describes the design and implementation of EFS in user space using faster cryptographic algorithms on UNIX Operating system. Implementing EFS in user space makes it portable & flexible; Kernel size will also not increase resulting in more reliable & efficient Operating System. Encryption techniques for file system level encryption are described, and general issues of cryptographic system interfaces to support routine secure computing are discussed.**

## I. INTRODUCTION

Encrypted File System is an interface that ensures the user that the data stored on the hard disk is secure and cannot be hacked by any other user without the permission of the owner. It ensures that the original data doesn't reside on the hard disk in the normal or the plaintext form, but it should always been stored in encrypted form which cannot be understood by the intruder. As in our current file system, it is normally stored in plaintext on the hard disk. So, if someone hacks the data stored in hard disk then that person can easily access the data. But if the file is stored in an encrypted form on the hard disk, the hacking in such cases won't be so effective.

User should not be aware about the location where the encryption and decryption takes place. By using encryption and decryption methodologies, user can secure his data and store it on the hard disk in an unreadable format. Several recent incidents accentuate the need for a cohesive solution to the problem of storage security that protects data using strong cryptographic methods in both personal and organizational scenarios. This paper investigates the implications of cryptographic protection as a basic feature of the file system interface.

## II. RELATED WORK

There are many architectures and procedures available in these areas that have already been implemented. Very few of them are implemented in user space and most of them are in kernel space. Each one of them is having certain advantages and limitations. The crucial issues of both, systems level and user level cryptography are as mentioned below.

### A. ISSUES WITH USER LEVEL CRYPTOGRAPHY

The simplest approach for file encryption is available through a tool, such as the UNIX crypt program, that enciphers (or deciphers) a file or data stream with a specified key. Depending on the particular software, the program may or may not be automatically delete the clear text while encrypting and such programs can usually be used as cryptographic "filters" in a command pipeline.

Another approach is integrated encryption in application software, where each program which has to manipulate sensitive data has built in cryptographic facilities. For example, a text editor could ask for a key when a file is opened and automatically encrypt and decrypt the file's data as they are written and read. All those applications that will be operated on the same data must, include the same encryption engine. An encryption filter, such as crypt, might also be provided to allow data to be imported into and exported out of other software.

Unfortunately, neither approach is entirely satisfactory in terms of security, generality or convenience. The former approach, which allows great flexibility in its application, invites mistakes; the user could inadvertently fail to encrypt a file, leaving it in the clear, or could forget to delete the clear text version after encryption. The manual nature of the encryption and the need to supply the key several times whenever a file is used makes encryption too cumbersome. More seriously, even when used properly, manual encryption programs open a window of vulnerability while the file is in clear form. It is almost impossible to avoid occasionally storing clear text on the disk and, in the case of remote file servers, sending it over the network. Some applications simply expect to be able to read and write ordinary files. In the application based approach, each program must have built in encryption functionality. Although encryption takes place automatically, the user still must supply a key to each application, typically when it is invoked or when a file is first opened. Software without encryption capability cannot operate on secure data without the use of a separate encryption program, making it hard to avoid all the problems outlined in the previous text. Furthermore, rather than being confined to a single program, encryption is spread among multiple applications, each of which must be trusted to interoperate securely and correctly with the others. A single poorly designed component can introduce a significant and difficult to detect window of vulnerability. Changing the encryption algorithm entails modification of every program that uses it, creating many opportunities of implementation errors. Finally, multiple copies of user level cryptographic code can introduce a significant performance penalty. [1]

## B. ISSUES WITH SYSTEM LEVEL CRYPTOGRAPHY

One way to avoid many of the pitfalls of user level encryption is to make cryptographic services a basic part of the underlying system. In designing such a system, it is important to identify exactly what is to be trusted with clear text and what requires cryptographic protection i.e. we must understand what components of the system are vulnerable to compromise.

For files, we are usually interested in protecting the physical media on which sensitive data are stored. This includes online disks as well as backup copies, which may persist long after the online versions have been deleted. In distributed file server based systems, it is often also desirable to protect the network connection between client and server since these links may be very easy for interception attack [1].

Physical media can be protected by specialized hardware. Disk controllers are commercially available with embedded encryption hardware that can be used to encipher entire disks or individual file blocks with a specified key. Once the key will be provided to the controller hardware, encryption will be completely transparent. This approach has a number of limitations for general uses. The granularity of encryption keys must be compatible with the hardware; often, the entire disk must be thought of as a single protected entity. It is difficult to share resources among users who are not willing to trust one another with the same key. Obviously, this approach is only applicable when the required hardware is available.

Network connections between client machines and file servers can be protected with end-to-end encryption. Specialized hardware may be employed for this purpose, depending on the particular network involved, or it may be implemented in software. However, all networks does not support encryption and among those that do, all system vendors does not supply working implementations of encryption as a standard product.[8]

Even when the various problems with media and network level encryption are ignored, the combination of the two approaches may not be adequate for the protection of data in modern distributed systems. In particular, even though clear text may never be stored on a disk or sent "over the wire", sensitive data can be leaked if the file server itself is compromised. At some point file server must maintain, the keys used to encipher both the disk and the network. Even if the server can be completely trusted, direct media encryption on top of network encryption has a number of shortcomings from the point of view of efficient distributed system design [2].

Further, the alternative approach taken by the Encrypted File System (EFS) has been mentioned. EFS pushes file encryption entirely into the client file system interface, and therefore does not suffer from many of the difficulties inherent in user level and disk and network based system level encryption.

## III. CRYPTOGRAPHIC SERVICES IN THE FILE SYSTEM

The main focus of EFS is to identify the location in a system, where file encryption will be performed. If it is at too low level, then trusts in components are removed from the user's control. If it is too close to the user, frequent human interaction may lead to error.

### A. DESIGN GOALS

EFS occupy something of a middle ground between low level and user level cryptography. It aims to protect exactly those aspects of file storage that are vulnerable to attack in a way that is convenient enough to use routinely. In particular, we will be guided by the following specific goals:

Key management scheme: The sensitive information of the file in encrypted file system is access by the key which is taken as the input from the user as a passphrase. This key is used to encrypt the content of the file and also help in returning the original content of the file. There must be a way to get this key from the user. It is taken in the form of passphrase. It is the most crucial input to the encrypted file system on which the whole security of the system relies.

Transparent access semantics: Encrypted files should support the same access methods available on the underlying storage system. All system calls should work normally, and it should be possible to compile and execute in a completely encrypted environment.

Transparent performance: Encryption algorithms are computationally intensive, but the overhead should not be too high so as to discourage the use of encrypted file system in the real scenario.

Security of file contents: The file contents should be secure effectively so that no other person gets to know about the data without the knowledge of key, the structural data should also be protected e.g. Information in the header & footer of the file should not generate same cipher text.

Natural key granularity: It should be easy to protect related files under the same key, and it should be easy to create new keys for other files. The UNIX directory structure is a flexible, natural way to group files.

Compatibility with underlying system services: The files and directory generated by the encrypted file system should behave normally and should be manageable as the normal file in the file system.

Portability: The encrypted file system should use the available functionality and features. The files and directory should be seen normally whenever the key is supplied to the file system.

Scalability: The encryption engine should not place an unusual load on any shared component of the system. File servers in particular should not be required to perform any special additional processing for clients who require cryptographic protection.

Compatibility with future technologies: Several emerging technologies have potential applicability for protecting data. In particular, keys could be contained in or managed by "smart cards" that would remain in the physical possession of authorized users. An encryption system should support, but not require, novel hardware of this sort.

**B. EFS FUNCTIONALITY AND USER INTERFACE**

Encrypted File System interacts with standard UNIX file system through system calls and treats all the files in same manner, irrespective of file being encrypted or normal file of standard file system. It prevents user from entering the same key several times. The EFS attaches a key to a directory and all the files within that directory are automatically encrypted. When this directory is attached to the Encrypted File System directory, then all the operations on the file can be executed. The files are automatically decrypted when they are read and are encrypted when write operation is performed. No modifications are required on the file system on which encrypted files are stored.

EFS provides "Virtual File System" on client's machine typically mounted on /crypt, through which user access their encrypted files. All the files are stored in the encrypted form and with the encrypted path name in associated directory. These files are not visible to the user until they attach the associated directory to the /crypt of the EFS. The underlying encrypted directories can reside on any accessible file system, including remote file servers such as Sun NFS [6]. No space is required to be pre-allocated for EFS directories and user controls EFS through commands like create, attach, detach etc.

To use Encrypted File System user has to create a directory and EFS by issuing command emkdir, with this key associate a passphrase i.e. key which is used by EFS to encrypt all the files within that directory. The passphrase should be at least 16 characters long. For instance, it can be "This is Encrypted File System". The emkdir works same as mkdir of DOS, but here we have to give passphrase in order to make it secure.

Eg **$ emkdir/user/jas/efs**   (name of the encrypted directory) (user enters passphrase, which does not echo) (same phrase entered again to prevent errors)

In order to use the files in the directory in the normal form, we have to supply key to the EFS. This is achieved by attach command. It takes three parameters.

1. Passphrase

2. Name of directory created

3. New name of directory

**$ attach /user/jas/efs aks** Key: (same key used in the cmkdir command)

If the key is supplied correctly, the user "sees" /crypt/aks as a normal directory; all standard operations (creating, reading, writing, compiling, executing, cd, mkdir, etc.) works as expected. The actual files are stored under /user/jas/efs, which would not ordinarily be used directly. Access to attached directories is controlled by restricting the virtual directories created under /crypt using the standard UNIX file protection mechanism. Only the user who issued the attach command is permitted to see or use the clear text files. This is based on the uid of the user; an attacker who can obtain access to a client machine and compromise a user account can use any of that user's currently attached directories. If this is a concern, the attached name can be marked obscure, which prevents it from appearing in a listing of /crypt. When an attach is made obscure, the attacker must guess its current name, which can be randomly chosen by the real user. Of course, attackers who can become the "super user" on the client machine can thwart any protection scheme, including this; such an intruder has access to the entire address space of the kernel and can read (or modify) any data anywhere in the system.

In order to remove the directory as files from /crypt we will use the command detach which removes the entry from the /crypt of EFS. File names are encrypted and encoded in an ASCII representation of their binary encrypted value padded out to the cipher block of size eight bytes.

**$ detach aks**

Some data are not protected. File sizes, access times, and the structure of the directory hierarchy are all kept in the clear text. (Symbolic link pointers are, however, encrypted.) This makes EFS vulnerable to traffic analysis from both real time observation and snapshots of the underlying files; whether this is acceptable must be evaluated for each application.

## IV. FILE ENCRYPTION METHODOLOGY

EFS use Advance Encryption Standard (AES) [4] to encrypt the file data. There are various modes of AES; one of it is Electronic Code Book (ECB) mode. [5] In which each eight byte block is encrypted individually. The main shortcoming of ECB is that it will produce same cipher text for the identical plain text block. It will help the cryptanalyst to find structural similarity of the data and help to decrypt the text easily.

Other modes of AES are various chaining ciphers. These modes help in reducing the shortcoming of ECB mode. It helps in overcoming the problem of structural analysis. But the problem with this mode is that, the random access of file becomes difficult due to reason that all the blocks are dependent on the cipher preceding block.

Since 56 bit key is vulnerable to exhaustive search of the key space. To remove this problem we will use multiple AES to provide more security to data. To remove both the shortcomings of random access and structural analysis, we had used both the modes of AES. The 128 bit supplied key is crunched into two halves of 56 bit key. Now the first 56 bit is used to calculate the long initial block with AES with ECB mode. Now whenever a file is to be written, it is XOR'ed with the initial block and then encrypted by the second key using AES with ECB mode. When reading, the cipher is reversed in the obvious manner, first decrypt in ECB mode then XOR it with initial block.

This method helps us to overcome both the problems of random access and structural analysis. It is clear that the protection against attack is at least as strong as a single AES pass in ECB mode and may be as strong as two passes with AES stream mode cipher. It is likely that the scheme is weakened, in such situation the attacker might be able to search for the two AES sub keys independently. If there are several known plaintext file encrypted with same key. In the chaining mode, as far as the Initialization Vector (IV) is different, the cipher text of identical block will be different. For this purpose we can attach IV to the beginning of file for maintaining atomicity.

Encryption of path name components uses a similar scheme with the addition that the higher order bit of clear text name are set to a simple checksum computed over the entire name string.

It is important to emphasize that EFS protects data only in the context of the file system. It is not, in itself, a complete, general purpose cryptographic security system. Once bits have been returned to a user program, they are beyond the reach of EFS's protection. This means that even with EFS, sensitive data might be written to a paging device when a program is swapped out or revealed in a trace of a program's address space. Systems where the paging device is on a remote file system are especially vulnerable to this sort of attack. (It is theoretically possible to use EFS as a paging file system, although the current implementation does not readily support this in practice.) .It should also be taken into consideration that EFS does not protect the links between users and the client machines on which EFS runs; users connected via networked terminals remain vulnerable if these links are not otherwise secured.

## A. PROPOSED ARCHITECTURE

The EFS prototype has been implemented entirely at user level, communicating with the Unix kernel via the NFS interface. Each client machine runs a special NFS server, efsd (EFS Daemon), on its localhost interface, that interprets EFS file system requests. At boot time, the system invokes efsd and issues an NFS mount of its localhost interface on the EFS directory (/crypt) to start EFS. (To allow the client to also work as a regular NFS server, EFS runs on a different port number from standard NFS.) The NFS protocol is designed for remote file servers, and so assumes that the file system is very loosely coupled to the client (even though, in EFS's case, they are actually the same machine)[6]. The client kernel communicates with the file system through remote procedure calls (RPCs) that implement various file system related primitives (read, write, etc.). The server is stateless, in that, it is not required to maintain any state data between individual client calls. All communication is initiated by the client, and the server can simply process each RPC as it is received and then wait for the next. Most of the complexity of an NFS implementation is in the generic client side of the interface, and it is therefore often possible to implement new file system services entirely by adding a simple NFS server.

*efsd* is implemented as an RPC server for an extended version of the NFS protocol. Additional RPCs attach, detach, and otherwise control encrypted directories. Initially, the root of the EFS file system appears as an empty directory. The attach command sends an RPC to efsd with arguments containing the full path name of a directory, the name of the "attach point", and the key. If the key is correct, cfsd computes the cryptographic mask (described in the previous section) and creates an entry in its root directory under the specified attach point name. The attach point entry appears as a directory owned by the user who issued the attach request, with a protection mode to prevent others from seeing its contents.

Encryption of pathname components uses a similar scheme, with the addition that the high order bits of the clear text name (which are normally zero) are set to a simple checksum computed over the entire name string. This frustrates structural analysis of long names that differ only in the last few characters. The same method is used to encrypt symbolic link pointers.

For each encrypted file accessed through an attach point, efsd generates a unique file handle that is used by the client NFS interface to refer to the file. For each attach point, the EFS daemon maintains a table of handles and their corresponding underlying encrypted names. When a read or write operation occurs, the handle is used as an index into this table to find the underlying file name. efsd uses regular Unix system calls to read and write the file contents, which are encrypted before writing and decrypted after reading, as appropriate. To avoid repeated open and close calls, efsd also maintains a small cache of file descriptors for files on which there have been recent operations. Directory and symbolic link operations, such as readdir, readlink, and lookup are similarly translated into appropriate system calls and encrypted and decrypted as needed. To prevent intruders from issuing RPC calls to EFS directly (and thereby thwarting the protection mechanism), efsd only accepts RPCs that originate from a privileged port on the local machine. Responses to the RPCs are also returned only to the localhost port, and file handles include a cryptographic component selected at attach time to prevent an attacker on a different machine from spoofing one side of a transaction with the server.

It is instructive to compare the flow of data under EFS with that taken under the standard, unencrypted file system interface. *Figure 1* shows the architecture of the interfaces between an application program and the ordinary Sun "vnode based" Unix file system [3]. Each arrow between boxes represents data crossing a kernel, hardware, or network boundary; the diagram shows that data written from an application are first copied to the kernel and then to the (local or remote) file system. *Figure 2* shows the architecture of the user level EFS prototype. Data are copied several extra times; from the application, to the kernel, to the EFS daemon, back to the kernel, and finally to the underlying file system. Since EFS uses user level system calls to communicate with the underlying file system, each file is cached twice, once by EFS in clear form and once by the underlying system in encrypted form. This effectively reduces the available file buffer cache space by a factor of two.

The architecture described above helps in analyzing the efficiency of the algorithm. To analyze an algorithm is to determine the amount of resources (such as time and storage) necessary to execute it. Most algorithms are designed to work with inputs of arbitrary length. Usually the efficiency or complexity of an algorithm is stated as a function relating the input length to the number of steps (time complexity) or storage locations (space complexity).

**Figure 1:** The Architecture of normal V-node file system in Unix

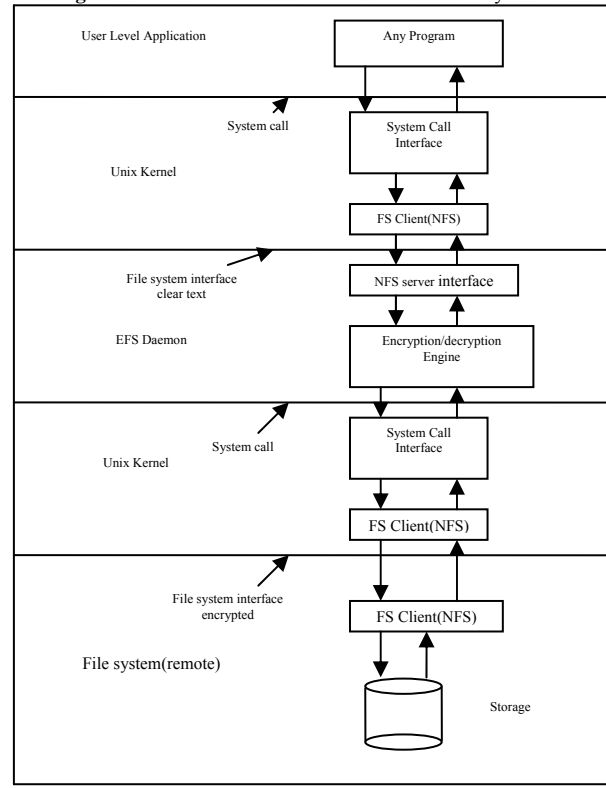

**Figure 2:** The Architecture of the user level EFS prototype

$$f(n) = \sum_{i=0}^{n-1}((\sum_{i+1}^{n-1}1 + \frac{1}{2}) + 3)$$

$$= \sum_{i=0}^{n-1}((n-i-1)\frac{3}{2} + 3)$$

$$= \frac{3}{2}\sum_{i=0}^{n-1}(n-i-1) + 3n$$

$$= \frac{3}{2}((n-1) + (n-2) + \cdots + 1) + 3n$$

$$= \frac{3}{2}n(n-1) + 3n$$

$$= \frac{3}{2}n^2 + \frac{3}{2}n$$

Therefore this algorithm has complexity $O(n^2)$. Its an important issue that while considering algorithms, one often measures complexity via the number of comparisons that occur, ignoring things such as assignments, etc. It will be suitable to keep track of any factors, in particular those which proceed with the dominating sub term. In the DES, the factor applied to the dominating sub term, namely $n^2$ was 3/2, and by coincidence, this was also the factor that came with the second term, namely n. It is obvious that an algorithm which is linear will perform better than a quadratic one, provided the size of the problem is large enough, but if it is known that the problem has a size of, say, at most 100 then a complexity of $(1/10)$ $n^2$ might be preferable to one of $1000000n$.

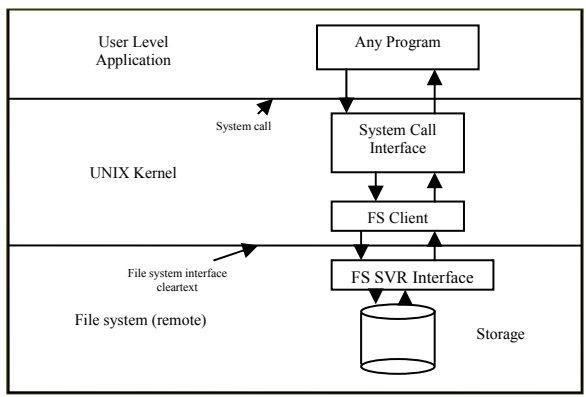

## V. RESULTS & PERFORMANCE EVALUATION

After implementing encrypted file system, it has been tried to find out the change in the space of file i.e. the variation of space of the encrypted file form the original file. For that some file of specific size has been taken and encrypted using the encrypted file system. The *Table-1* shows the variation of size of the original file when encrypted by EFS. The variation in size of encrypted file is approximately 2.5 times the size of the original file. As the encryption algorithm is computationally intensive, the computational time has been computed for the file of same size with EFS. The time taken by EFS to encrypt file of specific size is shown in *Table-2*. The final result is that both the time and space of the encrypted graph increase with the increase in the size of input file.

For the purpose of comparison standard utility of UNIX has been used i.e. *crypt*, which is used to encrypt and decrypt the file in the UNIX [7]. The same size of input files have been taken and the size of output file have been analyzed to get the variation in time and space and compared it with the time and space variation of encrypted file system. The variation in size is shown in *Table-3* and the variation in the time is shown in the *Table-4*.

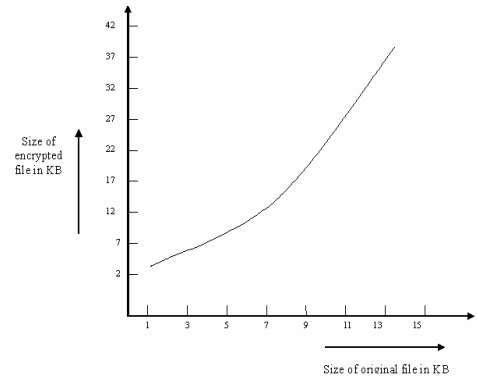

**Figure 3**: Variation in size of original file

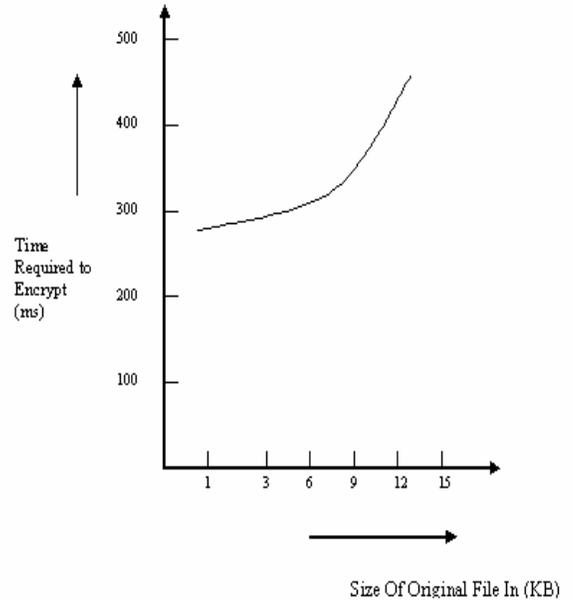

**Figure 4:** Variation in time according to size

**Table-1:** Difference in size of original and encrypted file by EFS

| Size Of Original file | Size Of Encrypted File |
|---|---|
| 909 bytes | 2.3 KB |
| 3.6 KB | 9.3 KB |
| 9.5 KB | 24.5 KB |
| 10.7 KB | 27.5 KB |
| 15.6 KB | 39.9 KB |

**Table- 2:** Time taken by EFS to encrypt a file

| Size of Original File | Time Taken By Encrypted File System |
|---|---|
| 909 Bytes | 278 ms |
| 3.6 KB | 304 ms |
| 9.5 KB | 358 ms |
| 10.7 KB | 381 ms |
| 15.6 KB | 410 ms |

**Table -3:** Difference in size of original and encrypted file by crypt

| Size Of original file | Size Of Encrypted File By crypt |
|---|---|
| 909 Bytes | 1.6 KB |
| 3.6 KB | 4.5 KB |
| 9.5 KB | 14.5 KB |
| 10.7 KB | 16.3 KB |
| 15.6 KB | 22.5 KB |

performance on modern workstations appears to be within a range that allows its routine use. Obviously, it has shortcomings of a user-level NFS server based implementation.

The client file system interface appears to be the right place to protect file data. If we consider the other alternatives, encrypting at the application layer is inconvenient. Application based encryption leaves windows of vulnerability while files are in the clear or requires the exclusive use of special purpose applications on all encrypted files. At the disk level, the granularity of encryption may not match the user's security requirements, especially if different files are to be encrypted under different keys. Encrypting the network in distributed file systems, while useful in general against network based attack, does not protect the actual media and therefore still requires trust in the server not to disclose file data.

The main focus of EFS is to reduce the barriers that stand in the way of the effective and ubiquitous utilization of file encryption. This is especially relevant as physical media remains exposed to theft and unauthorized access. Whenever sensitive data is being handled, it should be the *modus operandi* that the data be encrypted at all times when it is not directly being accessed in an authorized manner by the applications. It can be implemented on modern operating systems without having to change the rest of the system. Better performance and stronger security may be achieved by running the file system in the kernel. Proposed model of EFS is more portable than other kernel based file systems because it interacts with a standard vnode interface, as the quick ports to Linux.

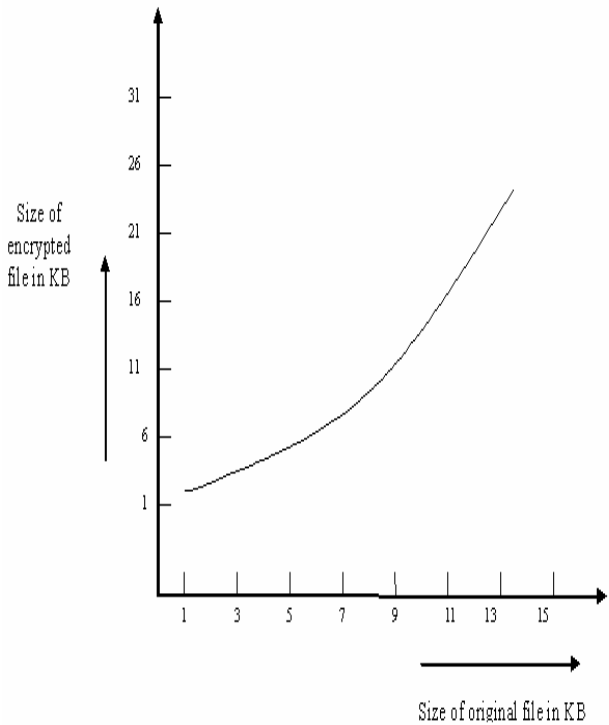

**Figure 5:** Variation of size of original file by Crypt

**Table-4:** Time taken by Crypt to encrypt a file

| Size Of Original File | Time Taken By crypt To Encrypt |
|---|---|
| 909 Bytes | ≈ 0 ns |
| 3.6 KB | ≈ 0 ns |
| 9.5 KB | 10000 ns |
| 10.7 KB | 10000 ns |
| 15.6 KB | 20000 ns |

## VI. CONCLUSIONS & FUTURE WORK

The proposed model of EFS provides a simple mechanism to protect data written to disks and sent to networked file servers. Although experience with proposed model of EFS is still limited to the research environment, rather

Network File System." *Proc. USENIX,* Summer, 1985.

**AUTHORS PROFILE**

**Dr. Shishir Kumar**

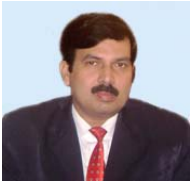

Dr. Shishir Kumar is currently working as Associate Professor and Head in Dept. of Computer Science & Engineering, Jaypee Institute of Engineering & Technology, Raghogarh Guna, India. He has completed his PhD in the area of Computer Science in 2005.He is having around 12 year teaching experience. He has published several papers in international/national referred journals and conferences. His area of Interest is Network Security & Image Processing.

**Mr. U.S. Rawat**

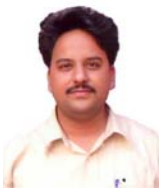

U.S. Rawat is currently working as Sr. Lecturer in Department of CSE, Jaypee Institute of Engineering & Technology, Raghogarh Guna, India. He received his B.E. in 1999 from Amravati University, Amravati ,India. He received his M.E. in 2003 from S.G.S.I.T.S, Indore, India. He is working towards his PhD from JUIT Waknagat, India. He is having eight years of teaching experience. His areas of interest are Information Systems & Network Security.